\documentclass[aps]{revtex4}
\usepackage{graphics}
\usepackage{epsfig}

\begin{document}

\title{Stability domains for time-delay feedback control with latency}

\author{Philipp H\"ovel}
\altaffiliation[Permanent address: ]
   {Institut f\"{u}r Theoretische Physik, TU Berlin, Hardenbergstrasse 36, D-10623 Berlin, Germany}
\affiliation{Department of Physics and Center for Nonlinear and Complex Systems, \\
         Duke University, Durham, NC 27708}
\author{Joshua E. S. Socolar}
\affiliation{Department of Physics and Center for Nonlinear and Complex Systems, \\
         Duke University, Durham, NC 27708}

\date{\today}

\begin{abstract} 
  We generalize a known analytical method for determining the
  stability of periodic orbits controlled by time-delay feedback
  methods when latencies associated with the generation and injection
  of the feedback signal cannot be ignored.  We discuss the case of
  extended time-delay autosynchronization (ETDAS) and show that
  nontrivial qualitative features of the domain of control observed in
  experiments can be explained by taking into account the effects of
  both the unstable eigenmode and a single stable eigenmode in the
  Floquet theory.
\end{abstract}

\maketitle

\section{Introduction}
Throughout the last decade the use of time delayed signals for
controlling unstable periodic orbits (UPOs) has been a field of
increasing interest.  A method first introduced by Pyragas
\cite{PYR92}, known as ``time-delay autosynchronization'' (TDAS),
calculates the control force from the difference of the current state
to the state one period in the past. Socolar et al.\cite{SOC94} have
shown that this technique can be improved by using states further in
the past.  This generalization of TDAS is called ``extended
time-delay autosynchronisation'' (ETDAS).  One great advantage of
ETDAS over conventional feedback controller schemes is that it can be
applied to high frequency oscillators.  Since it employs a direct
comparison of continuous signals generated by the system itself the
only factors limiting the speed of the controller is the bandwidth of
the amplifiers and signal propagation times.  There is no need to
generate a reference signal independently.

Experiments on electronic oscillators have shown that ETDAS can be
effective \cite{SUK97}.  They also show, however, that the latency
time associated with signal propagation -- the time required to
compare the the current signal with its time delayed counterpart and
inject the feedback into the system -- can have important effects.
Just \cite{JUS99b} has shown how longer latency times decrease the
range of feedback gains over which control is achieved for simple
systems controlled by TDAS. Here we extend his analytic formalism,
which consists of a first-order perturbation theory in the gain, to
ETDAS and note some novel features of the behavior of the individual
Floquet multipliers.  We then show how the theory provides a
qualitative explanation of the shape of the domain of control observed
in the experiments of Sukow et al.\cite{SUK97}.

\section{System equations and perturbation theory}
In this section we review the formalism developed by Just,
generalizing to the case of ETDAS controllers. We use the same
notation as Bleich and Socolar \cite{BLE96} for the system equations
and feedback signal, and the notation of Just \cite{JUS99b} for the
Floquet theory parameters.

Consider a $N$-dimensional dynamical system defined by
\begin{equation}\label{eqn:system}
{\bf\dot{x}}(t)={\bf f}({\bf x}(t),\epsilon_0 + \epsilon(t)),
\end{equation}
where ${\bf x}(t)$ denotes a $N$-dimensional state vector,
$\epsilon_0$ is a parameter that can be modulated to achieve control,
and $\epsilon$ is a feedback signal. Assume that in the absence of
control, $\epsilon(t) = 0$, there exists an unstable periodic orbit
${\bf x}_0(t)$ with period $\tau$.  Let $\xi (t) = {\bf \hat n} \cdot
{\bf x}(t)$ be a component of ${\bf x}$ that can be continuously
measured. The ETDAS feedback signal can be written in several
equivalent forms:
\begin{eqnarray}\label{eqn:etdas}
\epsilon(t)
&=&\gamma \sum_{k=1}^{\infty}R^{k}\left[\xi(t-t_l-k\tau)-\xi(t-t_l-(k+1)\tau)\right]
\nonumber\\
&=&\gamma\left(\xi(t-t_l)-(1-R)\sum_{k=1}^{\infty}R^{k-1}\xi(t-t_l-k\tau)\right)\nonumber \\
&=&\gamma \left(\xi(t-t_l)-\xi(t-t_l-\tau)\right)+R \epsilon(t-\tau),
\end{eqnarray}
where $\gamma$ (the feedback gain) and $R \in (-1,1)$ are real
parameters, and $t_l$ is the latency time. The control force vanishes
if the UPO is stabilized, since $\xi(t-k\tau) = \xi(t)$ for all $t$
and any integer $k$. TDAS corresponds to the special case $R=0$.  The
last form is the basis for simple implementation of ETDAS in
experiments.

In order to determine whether the controlled orbit is stable,
Equation~(\ref{eqn:system}) is linearized around the UPO. Defining
${\bf y}(t) = {\bf x}(t) - {\bf x}_0(t)$, we have
\begin{eqnarray}\label{eqn:linear}
{\bf \dot y}(t) &=& {\bf J}(t)\cdot {\bf y}(t) + \epsilon(t) \left.\frac{\partial
{\bf f}}{\partial \epsilon}\right|_{{\bf x}_0(t),\epsilon =0}\\ 
&=& {\bf J}(t)\cdot {\bf y}(t) + \gamma {\bf M}(t)\cdot \left({\bf
y}(t-t_l)-(1-R)\sum_{k=1}^{\infty}R^{k-1}{\bf y}(t-t_l-k\tau)\right),\nonumber
\end{eqnarray}
where ${\bf J}(t)=\left.{\bf J}(t)\right|_{{\bf x}_0(t),\epsilon =0}$
denotes the Jacobian matrix of the uncontrolled system, and ${\bf M}(t)
=\left.\frac{\partial {\bf f}}{\partial \epsilon}\right|_{{\bf
x}_0(t),\epsilon=0} \otimes {\bf \hat n}$ is a $N \times N$ matrix
containing all information about the control force.

Since ${\bf J}(t)$ and ${\bf M}(t)$ are both periodic with period
$\tau$, Floquet theory ensures that ${\bf y}(t)$ can be written as
\begin{equation}\label{eqn:floquet}
{\bf y}(t) = \sum_{m=0}^{\infty}\sum_{n=1}^{N}c_m^{(n)} e^{(\Lambda_m^{(n)} + i
\Omega_m^{(n)})t} {\bf p}_m^{(n)}(t),
\end{equation}
where ${\bf p}_m^{(n)}(t)$ is a periodic function with period $\tau$:
\begin{equation}
{\bf p}_m^{(n)}(t) =  {\bf p}_m^{(n)}(t+\tau).
\end{equation}
The factorization of the sum into a double sum is done for convenience
in the discussions below.  In the absence of control, i.e. the absence
of time delay terms, there are $N$ eigenmodes of the system, indexed
by $n$.  When control is turned on, each of these gives rise to a
countably infinite set of eigenmodes indexed by $m$.  For each set,
there is one eigenvalue that begins at the original value
$\lambda^{(n)}+i\omega^{(n)}$ for $\gamma=0$ and varies as $\gamma$ is
increased.  The remaining members all have eigenvalues that approach
either $\log|R|/\tau$ or $-\infty$ as $\gamma$ approaches $0$.

$\Lambda_m^{(n)}$ and $\Omega_m^{(n)}$ are the real and imaginary
parts of the Floquet exponent corresponding to the eigenmode ${\bf
  p}_m^{(n)}(t)$.  Inserting Equation~(\ref{eqn:floquet}) into
Equation~(\ref{eqn:linear}) will lead to conditions that must be
satisfied by $\Lambda_m^{(n)}$ and $\Omega_m^{(n)}$.  The system is
linearly stable if and only if all $\Lambda_m^{(n)}$ that satisfy
these conditions are negative.  Equations~(\ref{eqn:linear})
and~(\ref{eqn:floquet}) yield the following equation for each of the
modes ${\bf p}_m^{(n)}(t)$, where we drop the subscript $m$ and
superscript $n$ for notational convenience:
\begin{equation}  \label{eqn:eigenvalue}
(\Lambda + i \Omega){\bf p}(t)+{\bf \dot p}(t)={\bf J}(t){\bf p}(t)+\gamma {\bf M}(t)e^{-(\Lambda + i \Omega)t_l}\frac {1-e^{-(\Lambda + i
\Omega)\tau}}{1-R \,e^{-(\Lambda + i \Omega)\tau}} {\bf p}(t-t_l).
\end{equation}

This equation is equivalent to
\begin{eqnarray}\label{eqn:pdot}
(\Lambda + i \Omega) {\bf p}(t;\kappa)+{\bf \dot p}(t;\kappa)= \left[{\bf J}(t)+
\kappa {\bf M}(t)\cdot {\bf W}(t,-t_l) \right]\cdot {\bf p}(t;\kappa),
\end{eqnarray}
where ${\bf W}(t,\Delta t)$ is the propagator defined by ${\bf
p}(t;\kappa) = {\bf W}(t,\Delta t)\cdot{\bf p}(t-\Delta t;\kappa)$ and
\begin{equation}
\kappa \equiv \gamma e^{-(\Lambda + i\Omega)t_l}\frac {1-e^{-(\Lambda + i \Omega)\tau}}{1-R\,e^{-(\Lambda + i \Omega)\tau}}.
\end{equation}
Since $\kappa$ is proportional to $\gamma$, it can be loosely thought
of as a measure of the strength of the control gain.  One must keep in
mind, however, that the value of $\kappa$ is ultimately determined by
the solutions for the exponent $\Lambda +i\Omega$.  Note also that the
Floquet eigenmodes ${\bf p}(t;\kappa)$ themselves depend on $\Lambda
+i\Omega$ through $\kappa$, making for a nontrivial modification of
the usual eigenvalue problem.

An expression taking the effects of control into account can be
derived by perturbation theory.  Equation~(\ref{eqn:pdot}) can be
written as
\begin{equation}\label{eqn:perturbation}
(\Lambda + i \Omega){\bf p}(t;\kappa)=\left(-\frac{d}{dt}+{\bf J}(t)+
\kappa {\bf M}(t)\cdot {\bf W}(t,-t_l) \right)\cdot{\bf p}(t;\kappa)(t).
\end{equation}
We regard $-\frac{d}{dt}+{\bf J}(t)$ as an operator with known
eigenvalues $\lambda^{(n)} = i \omega^{(n)}$ and consider $\kappa {\bf
M}(t)\cdot {\bf W}(t,t_l)$ to be a perturbation, a technique familiar
from quantum mechanics. The effects of the controller on the Floquet
exponents can be expanded in powers of $\kappa$ as
\begin{equation}\label{eqn:first-order}
\Lambda + i \Omega = \lambda + i \omega + \chi(t_l) \kappa+o(\kappa^2),
\end{equation}
where the coefficient $\chi(t_l)$ is a complex valued function.  Any
effects of interactions between the Floquet modes enter only at
$\kappa^2$.  This is why it is convenient to index the modes by $n$
and $m$ for the purposes of first-order perturbation theory.

As mentioned above, Equation~(\ref{eqn:first-order}) has an infinite
number of solutions for $\Lambda$ and $\Omega$ which approach
$\log|R|/ \tau$ \cite{PYR01,BEC02} or minus infinity as the feedback
gain $\gamma$ goes to zero.  This behavior arises from the essential
singularity at $\Lambda = \log|R| / \tau$ and the divergence at
$\Lambda = -\infty$, the latter arising only due to the nonzero
latency time.

Just \cite{JUS99b} has shown how the coefficient $\chi(t_l)$ can be
calculated. Let ${\bf u}(t) = {\bf p}(t;\kappa =0)$ and ${\bf v}^*(t)$
be the right and left Floquet eigenmode in the absence of control,
$\kappa = 0$. Both are periodic with period $\tau$, i.e.  ${\bf
u}(t)={\bf u}(t+\tau)$ and ${\bf v}^*(t)={\bf v}^*(t+\tau)$.   Now let
${\bf \hat u}(t)=\exp(i \omega t){\bf u}(t)$ and ${\bf \hat
v}^*(t)=\exp(-i \omega t){\bf v}^*(t)$.  These satisfy the equations
\begin{eqnarray}\label{eqn:leftandrighthat}
\lambda {\bf \hat u}(t)+{\bf \dot {\hat u}}(t)&=&{\bf J}(t)\cdot {\bf \hat
u}(t)\nonumber\\ 
\lambda {\bf \hat v}^*(t)-{\bf \dot {\hat v}}^*(t)&=&{\bf \hat v}^*(t)\cdot
{\bf J}(t),
\end{eqnarray}
with the boundary conditons:
\begin{eqnarray}\label{eqn:hat}
{\bf \hat u} (t+\tau) &=& e^{i\omega \tau}{\bf \hat u} (t) \nonumber \\
{\bf \hat v}^*(t+\tau) &=& e^{-i\omega \tau}{\bf \hat v}^*(t).
\end{eqnarray}

The standard first-order perturbation theory result for the coefficient $\chi(t_l)$ is 
\begin{eqnarray}\label{eqn:chi2}
\chi(t_l)= e^{i\omega t_l} \rho(t_l),
\end{eqnarray}
where
\begin{equation} \label{eqn:rhohat}
\rho(t_l)=\frac{\int^\tau_0 {\bf \hat v}^*(t){\bf M}(t)\cdot {\bf W}(t,t_l){\bf \hat u}(t)dt}{\int^\tau_0 {\bf \hat
v}^*(t){\bf \hat u}(t)dt}.
\end{equation}
Now $\rho(t_l)$ depends on $t_l$ only through ${\bf W}$, and since
${\bf u}(t)$ is $\tau$-periodic, ${\bf W}(t,t_l+\tau){\bf u}(t)={\bf
  W}(t,t_l){\bf u}(t)$. Therefore, $\rho(t_l)$ has to satisfy
\begin{equation}\label{eqn:rho}
\rho (t_l+\tau) = e^{i\omega \tau} \rho (t_l).
\end{equation}
Inserting Equation~(\ref{eqn:chi2}) in Equation~(\ref{eqn:first-order}) and
neglecting second order terms yields
\begin{eqnarray}\label{eqn:rhoexp}
\Lambda + i \Omega = \lambda + i \omega + \rho(t_l) \gamma \,e^{-(\Lambda + i (\Omega-\omega)) t_l} \frac{1-e^{-(\Lambda + i \Omega) \tau}}{1-R \,e^{-(\Lambda
+ i \Omega) \tau}}.
\end{eqnarray}

Following the treatment of Just \cite{JUS99b} for the $R=0$ case, we
note that this expression can be simplified in the case of a so-called
flip orbit, where $\omega = \pi /\tau$.  Defining the frequency
deviation $\Delta\Omega = \Omega - \pi /\tau$,
Equation~(\ref{eqn:rhoexp}) can be rewritten as
\begin{equation}\label{eqn:flip}
\Lambda + i \Delta\Omega = \lambda + \rho(t_l) \gamma \,e^{-(\Lambda + i \Delta\Omega) t_l} \frac{1+e^{-(\Lambda + i \Delta\Omega) \tau}}{1+R \,e^{-(\Lambda
+ i \Delta\Omega) \tau}}.
\end{equation}

Moreover, since all coefficients in Equation~(\ref{eqn:leftandrighthat})
are real, and  for $\omega = \pi/\tau$ the boundary conditions of
Equation~(\ref{eqn:hat}) are invariant under complex conjugation, ${\bf
\hat u}(t)$ and ${\bf \hat v}^*(t)$ can always be chosen to be real
valued. From Equation~(\ref{eqn:rhohat}) it is then clear that
$\rho(t_l)$ is real and from Equation~(\ref{eqn:rho}) we have
$\rho(t_l)=-\rho(t_l+\tau)$. 

We emphasize that ${\bf \hat u}(t)$, ${\bf \hat v}^*(t)$, and
$\rho(t_l)$ are also real valued in the case of zero torsion, i.e.
$\omega = 0$.  In this case, $\chi(t_l)=\rho(t_l)$ and $\rho$ is
$\tau$-periodic: $\rho(t_l)=\rho(t_l+\tau)$. 
For torsion-free perturbations, Equation~(\ref{eqn:rhoexp}) becomes
\begin{eqnarray}
\Lambda + i \Omega = \lambda + \chi(t_l) \gamma \,e^{-(\Lambda + i
\Omega) t_l} \frac{1-e^{-(\Lambda + i \Omega) \tau}}{1-R \,e^{-(\Lambda + i
\Omega) \tau}}.
\end{eqnarray}
Although a nonzero torsion of all unstable eigenmodes is a necessary
condition for possible control, \cite{JUS97} the case of $\lambda < 0$
and $\omega = 0$ might be interesting because an initially stable
eigenmode can become unstable when the control force is applied and thus
limit the domain of control.

\section{Floquet exponents for fixed latency time}
To understand the solutions to Equation~(\ref{eqn:flip}), it is helpful to first
consider their behavior for fixed values of $g \equiv -\rho(t_l)\gamma$. A detailed
discussion of the effects of variations of $\rho$ with $t_l$ will be presented in
section~IV.
\begin{figure}[t] 
\epsfxsize=\linewidth
\epsfbox{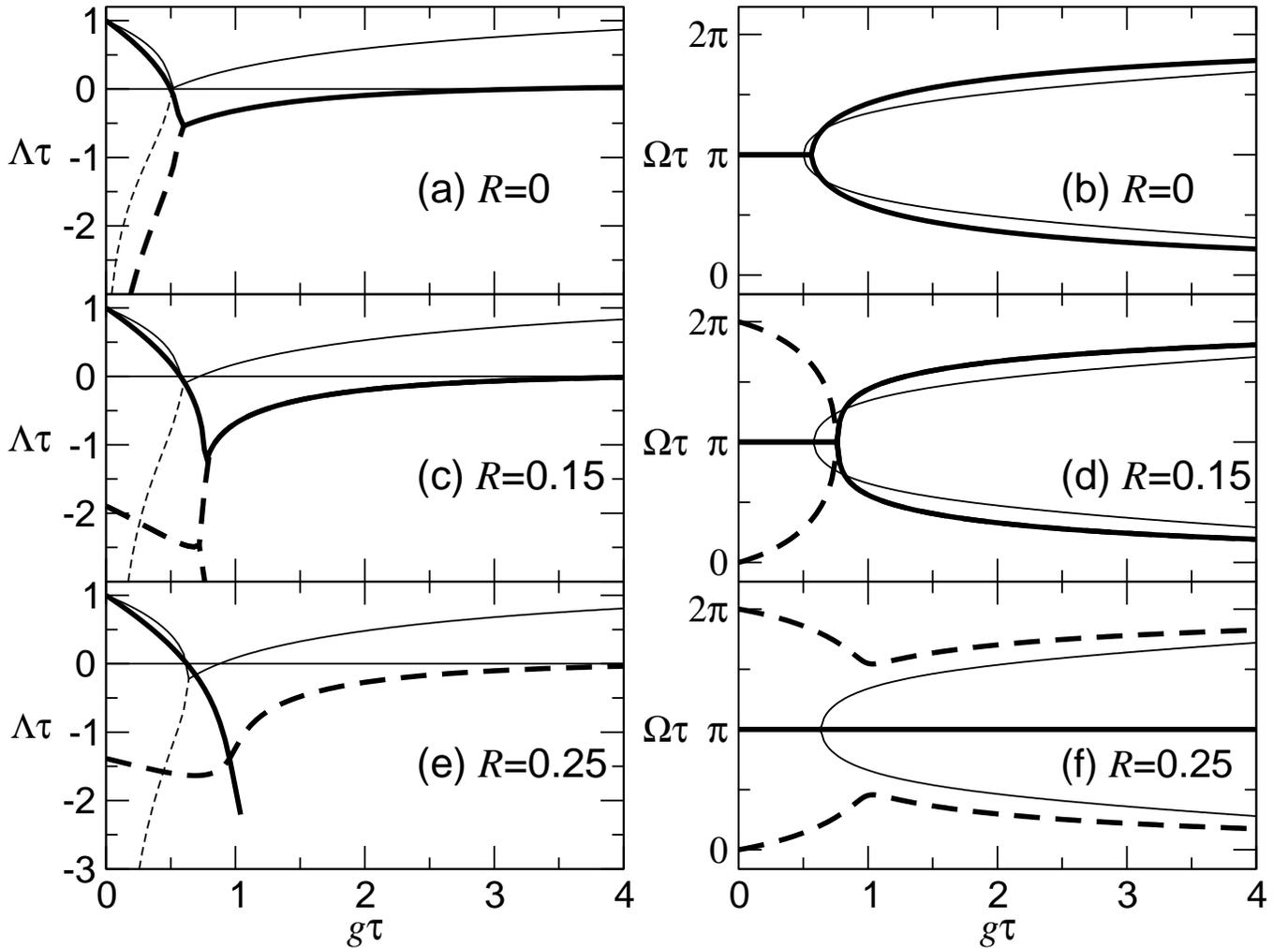}
\caption{\label{fig:exponent}  Real and imaginary part of the Floquet
  exponent versus $g$ for $\lambda \tau =1$. Thick lines correspond to a
  latency time $t_l = 0$, thin lines to $t_l = 0.5 \tau$ for different
  values of $R$: 0 (TDAS), 0.15, 0.25. The solid line represents the
  system's exponent, the dashed line represents the exponent created by
  the control scheme.}
\end{figure} 

The behavior of the real and imaginary parts of the Floquet exponent in
Equation~(\ref{eqn:flip}) can be seen in Figure~\ref{fig:exponent}. In
each panel, curves are shown for both $t_l = 0$ (thick lines) and $t_l =
\tau /2$ (thin lines). For $g=0$ (no control), the real part $\Lambda$ is
equal to $\lambda$. For increasing $g$ the value of $\Lambda$ decreases,
reaching $0$ at $g =\lambda (1+R) /2$, then changing its sign; thus the
orbit becomes stable. Further increase of $g$ usually leads to a
collision with an exponent created by the control scheme, forming a
complex conjugate pair (see Figure~\ref{fig:exponent}(b), (d), and (f),
except for the thick line in (f)). After the collision, $\Lambda$ then
begins to increase (see Figure~\ref{fig:exponent}(a), (c), and (e),
except for the thick line in (e)). For $g$ sufficiently large, $\Lambda$ becomes
positive again and control is lost.

Note that as $g$ increases from zero, an infinite number of solutions to
Equation~(\ref{eqn:flip}) emerge from $\log|R|/\tau$ as complex conjugate
pairs. \cite{PYR01,BEC02} In order to collide with the single exponent
coming from $(\lambda+i\omega)\tau$, one pair has to become real and
separate (see Figure~\ref{fig:exponent}(d)). If this does not happen, a
crossing of branches can occur. After the crossing, the complex conjugate
pair becomes the branch with largest $\Lambda$ and thus responsible for
the stability of the system (see  thick lines in
Figure~\ref{fig:exponent}(e) and~\ref{fig:exponent}(f)).

Figure~\ref{fig:exponent} illustrates that increasing the latency time
$t_l$ and/or decreasing $R$ leads to a smaller range of $g$ for successful
control. In fact, one can compute a maximum latency time $t_{max}$ for
which control can be achieved, which corresponds to the case where the
collision of the two branches occurs at $\Lambda =0$ as in the thin line
in Figure~\ref{fig:exponent}(a). Using the imaginary part of
Equation~(\ref{eqn:flip}) to eliminate the factor of $g$ in the real part
of that same equation and setting $\Lambda$ equal to zero, one has to
search for nontrivial solutions for $\Delta \Omega$. A
condition for the existence of such a solution is
\begin{equation}\label{eqn:tmax}
t_l \leq t_{max}=\frac{1}{\lambda}+\frac{\tau}{2}\,\frac{R-1}{R+1}.
\end{equation} 
This agrees with the result of \cite{JUS99b} for $R=0$. For larger
latency time, control is not possible to first-order in $\kappa$.
Equation~(\ref{eqn:tmax}) shows that negative $R$ decreases the maximum
latency time. We therefore focus on $R \geq 0$ from here on.

Another way to visualize the effects of the control scheme is to consider
how the Floquet {\em multipliers} (as opposed to exponents) evolve with
varying $g$. We consider again the case $\omega = \pi / \tau$. The Floquet
multipliers are defined by $\mu_m^{(n)} = e^{(\Lambda_m^{(n)} + i
\Delta\Omega_m^{(n)})\tau}$. Dropping the subscript $m$ and the
superscript $n$, Equation~(\ref{eqn:flip}) can be rewritten:
\begin{equation}\label{eqn:multiplier} \mu = \exp\left[\lambda \tau -
g\tau \left( \frac{\mu+1}{\mu+R} \right) \mu^{-t_l/\tau}\right]. 
\end{equation}
Stability is achieved if all multipliers satisfy $|\mu| < 1$; i.e.
the multipliers $\mu$ are located inside the unit circle in the
complex plane.
\begin{figure}[t] 
\epsfxsize=\linewidth
\epsfbox{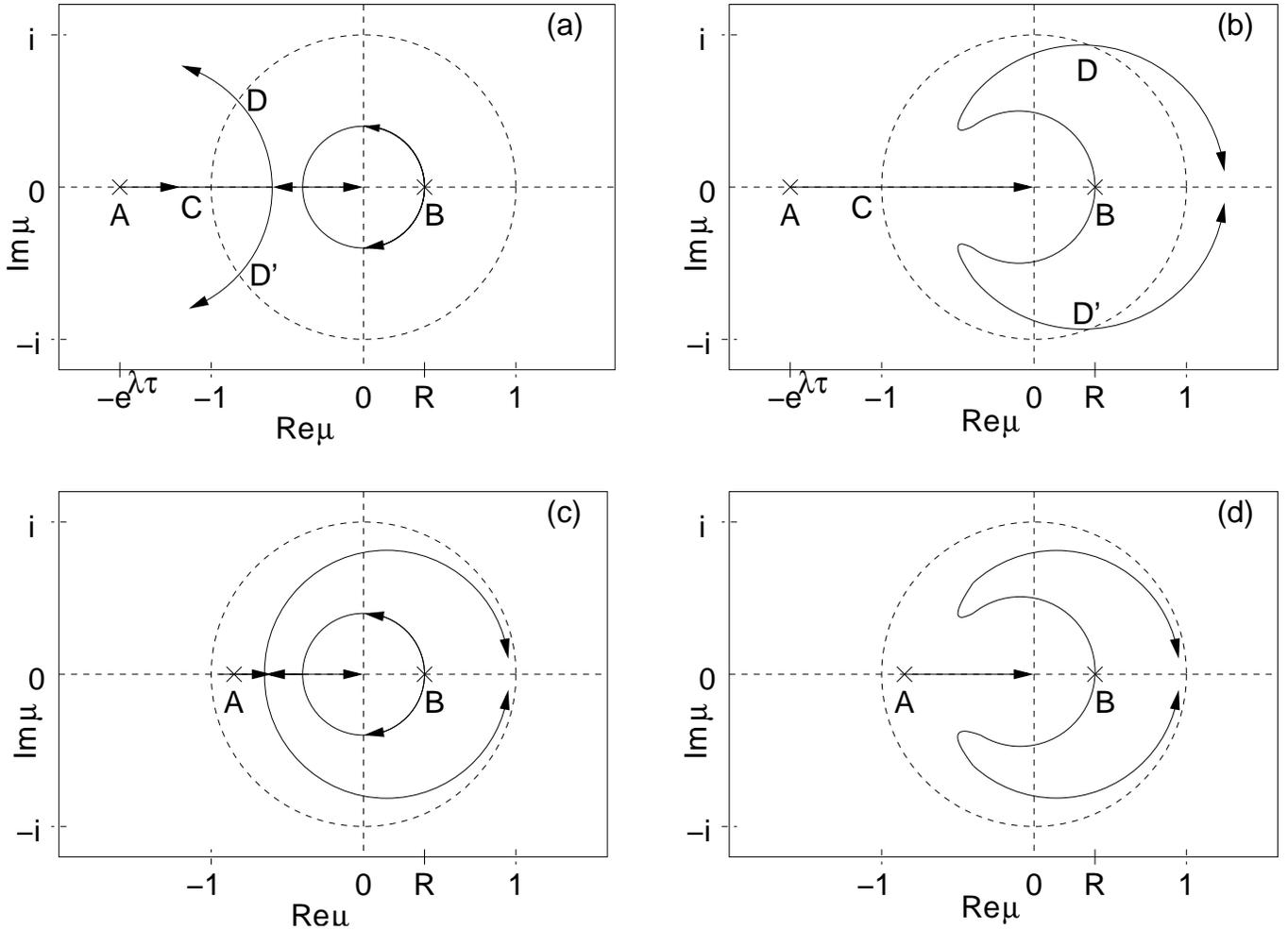}
\caption{\label{fig:multiplier}  Schematic behavior of the multiplier
  $\mu$ in the complex plane for $g > 0$ and $t_l = 0$ in the case $\omega
  = \pi / \tau$. Pictures (a) and (b) visualize the case of an initially
  unstable multiplier of the system, pictures (c) and (d) the case of an
  initially stable one. Point A indicates the uncontrolled multiplier of
  the system $\mu = -\exp(\lambda \tau)$, point B shows where the created
  multipliers start, point C is the point where control is obtained, and
  points D and D' are the points where control is lost. The arrows
  indicate the directions in which the multipliers move for increasing
  $g$.}
\end{figure}

Considering the real and imaginary part of Equation~(\ref{eqn:multiplier})
and numerically following the roots of each equation in the complex plane
(using Mathematica), we observe topologically different cases depending on
the signs of $\lambda$ and $g$ in the absence of latency, $t_l = 0$, as
illustrated in Figure~\ref{fig:multiplier}.
\begin{itemize}
\item $g>0;\lambda > 0$: (Figure~\ref{fig:multiplier}(a) and (b)) As $g$
  increases, the largest Floquet multiplier starts outside the unit
  circle at $-\exp(\lambda \tau)$ on the real axis indicated by point A,
  moves towards the unit circle, and eventually crosses it.  Meanwhile an
  infinite number of complex conjugate pairs spread out from $\mu = R$
  indicated by point B one of which will determine the stability range.
  Two scenarios are possible. One possibility is that one pair recombines
  on the real axis and one of the multipliers collides with the system's
  while the other approaches zero. After the collision they form a
  complex conjugate pair and cross the unit circle again (see
  Figure~\ref{fig:multiplier}(a)). The other possibilitiy is that the
  created pair becomes the largest multiplier, turns around and crosses
  the unit circle (see Figure~\ref{fig:multiplier}(b)). The first case
  corresponds to the thick lines in Figure~\ref{fig:exponent}(c) and (d),
  the second to Figure~\ref{fig:exponent}(e) and (f).
\item $g>0;\lambda < 0$:  (Figure~\ref{fig:multiplier}(c) and (d))
  Similar to the previous case the largest Floquet multiplier starts at
  $-\exp(\lambda \tau)$ on the real axis, this time inside the unit
  circle. It moves towards the origin for increasing $g$ and may either
  collide with a multiplier created by the control force as in
  Figure~\ref{fig:multiplier}(c), or continue towards zero while a
  complex conjugate pair becomes the largest multiplier as in
  Figure~\ref{fig:multiplier}(d). In both cases all multipliers stay
  inside the unit circle for increasing $g$; for $\lambda < 0$ the system
  is stable for all $g > 0$.
\item$g<0; \lambda > 0$: The largest multiplier starts at $-\exp(\lambda
  \tau)$ on the real axis and goes to $- \infty$, thus control is never
  successful. All other multipliers created by the control scheme stay
  inside the unit circle for decreasing $g$.
\item$g<0; \lambda < 0$: The largest multiplier starts inside the unit
  circle, crosses it, and goes to $- \infty$. More mulitpliers created by
  the control scheme cross the unit circle for further decrease of $g$. 
\end{itemize}

\section{Shapes of the domain of control in the $t_l$-$\gamma$ plane}
For a discussion of the domain of control in the $t_l$-$\gamma$ plane
let us consider first $\rho(t_l) = -1$. We will show later how the
coefficient $\rho(t_l)$ scales the domain at every value of the
latency time.

\begin{figure}[t]  
\epsfxsize=\linewidth
\epsfbox{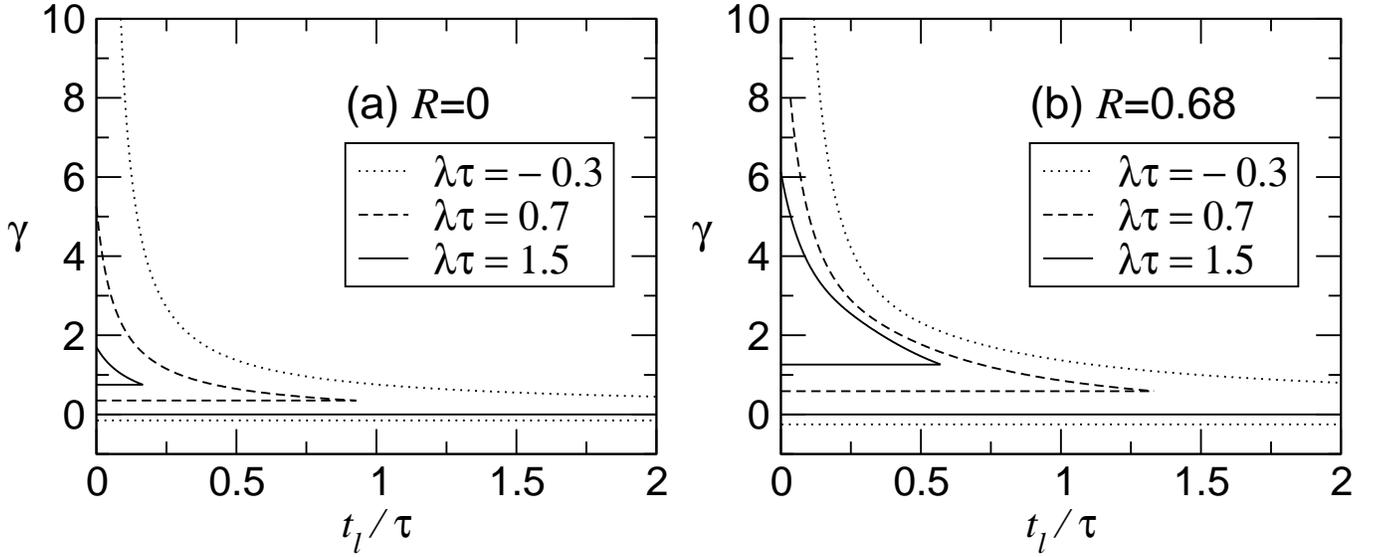}
\caption{\label{fig:rho=1}  Domain of control in the $t_l$-$\gamma$ plane
  for different values for $\lambda \tau$: -0.3, 0.7, 1.5 and
  $\rho(t_l)=-1$. Picture (a) shows the case of TDAS, $R=0$, and picture (b)
  the case of $R=0.68$. The branches indicate combinations for $t_l$ and
  $\gamma$ for which the real part of the Floquet exponent $\Lambda$
  changes sign and thus stability.}
\end{figure}

For each $\lambda$, the lower branch in Figure~\ref{fig:rho=1} is the
horizontal line $\gamma = \lambda (1+R)/2$, where there is a flip
instability associated with the real exponent originating from
$\lambda$ at $\gamma =0 $. The upper branch corresponds to a Hopf
bifurcation that can arise in two different ways: (1) the relevant
complex conjugate pair of exponents originates in a collision between
the branch associated with $\lambda$ and a real eigenvalue created by
the feedback scheme, as in Figure~\ref{fig:exponent}(a) or~(c); or (2)
the relevant complex conjugate pair originates at $\log|R|/\tau$, as
in the thick lines in Figure~\ref{fig:exponent}(d).  It can be seen
that increasing $R$ and decreasing $\lambda$ increases the domain of
control at fixed $t_l$ and increases $t_{max}$ (see
Equation~(\ref{eqn:tmax})). The upper and lower curves do not
intersect if the system is already stable, i.e.  $\lambda < 0$.
\begin{figure}[t]
\epsfxsize=\linewidth
\epsfbox{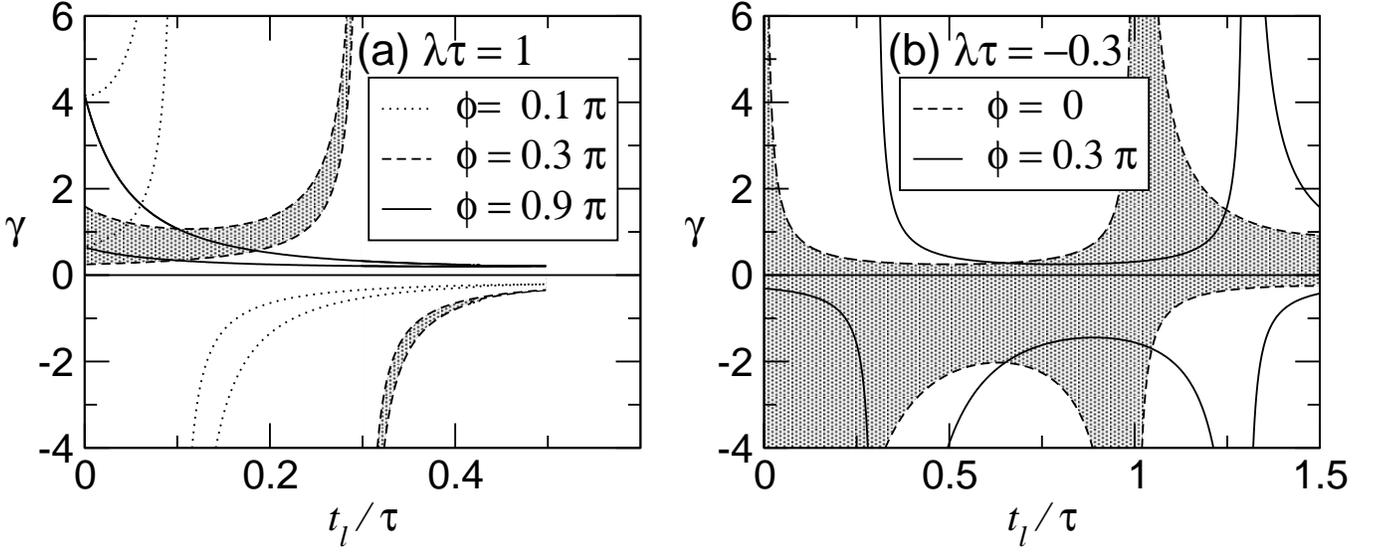}
\caption{\label{fig:rho=Asin}   Domain of control in the $t_l$-$\gamma$
  plane for $R = 0$ in case of $\rho(t_l) = A \, \sin(\frac{\pi}{\tau}
  t_l -\phi)$. Picture (a) shows the case of an UPO ($\lambda \tau=
  1$) with $A = 2.5$ for different values of the phase $\phi$: $0.1
  \pi, 0.3 \pi, 0.9 \pi$. Picture (b) shows the case of a stable orbit
  ($\lambda \tau= -0.3$) with $A = 0.6$ for $\phi$: $0, 0.7
  \pi$. The shaded regions show the domain of stability for $\phi =
  0.3 \pi$ and $\phi = 0$, respectively.}
\end{figure}

The actual domain of control is a distortion of Figure~\ref{fig:rho=1}
owing to the variation of $\rho$ with $t_l$. The distortion is simple to
compute, however, since changing the value of $\rho$ is entirely
equivalent to changing $\gamma$.  Thus the values of $\gamma$ on the
upper and lower curves at a particular value of $t_l$ are simply
multiplied by $-1/\rho(t_l)$. Note that these variations in $\rho$ cannot
change the {\em ratio} of the upper and lower values of $\gamma$.

For eigenmodes with $\omega = \pi / \tau$, the antiperiodicity of
$\rho(t_l)$ and the fact that $\rho (t_l)$ is real require that
$\rho(t_l)$ has at least one root in the interval $[0,\tau]$. We
assume for convenience a sinusoidal form $\rho(t_l) = A \,
\sin(\frac{\pi}{\tau} t_l -\phi)$.  Figure~\ref{fig:rho=Asin} shows
the domain of control for different values of the phase $\phi$.  In
each panel, the shading indicates the stable domain for one choice of
$\phi$.

If a root of $\rho(t_l)$ appears before the upper and lower branches
intersect, the domain of control will include a region with negative
feedback gain $\gamma$, as shown in Figure~\ref{fig:rho=Asin}(a).
Since the domain of control at fixed $t_l$ scales like $1/\rho$,
divergences appear at the values of the latency time for which
$\rho(t_l)$ vanishes. For $\lambda <0$, shown in
Figure~\ref{fig:rho=Asin}(b), the upper and lower branches still do
not intersect.
\begin{figure}[t]
\epsfxsize=0.45 \linewidth
\epsfbox{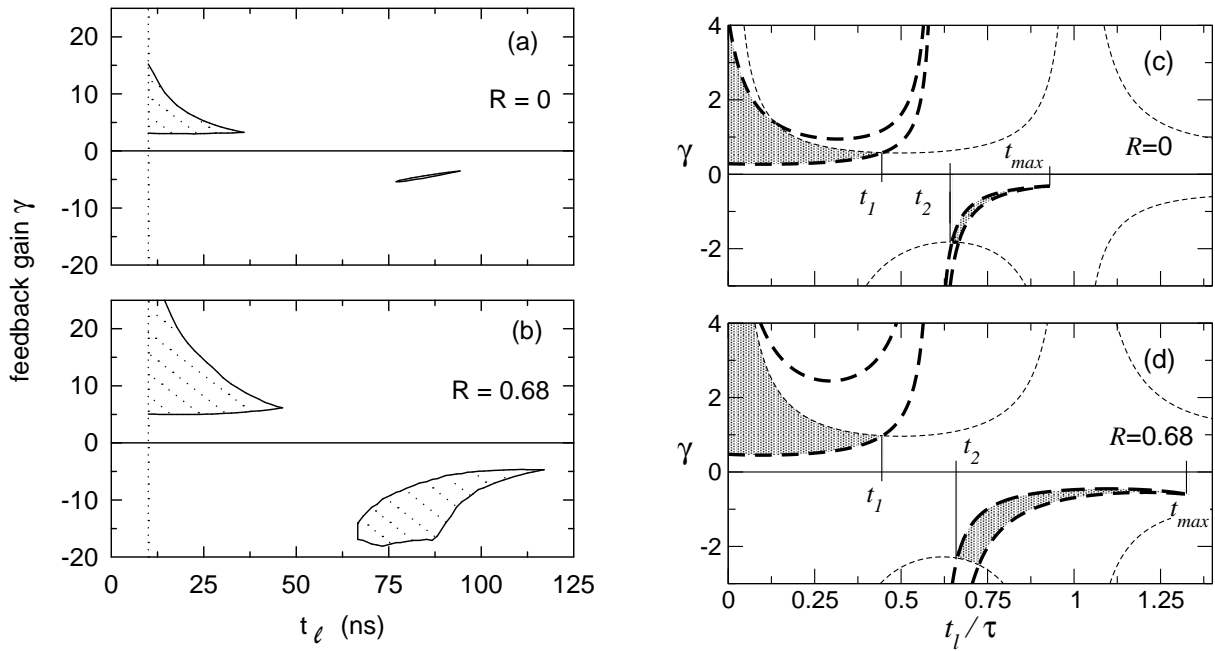}
\hspace{0.3in}
\epsfxsize=0.4 \linewidth
\epsfbox{figure5.eps}
\caption{\label{fig:regions}  
  Domain of control in the $t_l$-$\gamma$ plane.  (a) and (b):
  Experimental result from a diode resonator circuit investigated by
  Sukow et al.  This is a reproduction of Figure~18 from \cite{SUK97}.
  In this system, the period $\tau$ was equal to $100$ ns.  The scale
  for the gain is multiplied by an arbitrary factor.  (c) and (d):
  Domains determined from theory for two Floquet exponents with
  parameters chosen to reproduce as closely as possible the
  experimental results of (a) and (b).  The thick and thin dashed
  lines show the domains corresponding to the two different modes.
  (See Figure~\ref{fig:rho=Asin}.)  The shaded regions show the
  combined domain of control.  The parameter values are $\lambda_1
  \tau= 0.7$, $\lambda_2 \tau= -1.6$, $\omega_{1} = \omega_{2} =
  \pi/\tau$, $\rho_1(t_l)= 1.3 \sin (\frac{\pi}{\tau}t_l-0.6\pi)$ and
  $\rho_2(t_l)= 1.4 \sin (\frac{\pi}{\tau}t_l).$ The two panels show
  different values of the control parameter $R$.  The special latency
  times marked are used to determine the parameter values as explained
  in the text.}
\end{figure}

The case $\lambda <0$ can be important because it can reduce the
domain of control, as shown in Figure~\ref{fig:regions}(c) and (d),
when the uncontrolled system has exponents $\lambda_1 >0$ and
$\lambda_2 <0$.  Since second order terms are neglected in the
perturbation theory, the different Floquet modes of the system do not
interact with each other, so the effective domain of control is just
the intersection of the single domains for each exponent.

Domains similar to those shown in Figure~\ref{fig:regions}(c) and~(d)
have been observed in experiments on high speed diode resonator
circuits.  The analogous figure obtained from experiments is
reproduced here as Figure~\ref{fig:regions}(a) and~(b) to facilitate
comparison.  To construct the theoretical figure, parameter values
are adjusted to reproduce several features of the experimental
results.  From the experiments, three parameters are known: the
weighting parameter $R$, the largest Lyapunov exponent $\lambda_1$,
and, since the instability is a flip, the frequency $\omega_1 = \pi
/\tau$.  Equation~(\ref{eqn:tmax}) for the $\lambda_1$ mode then gives
an immediate prediction for $t_{max}$, the largest latency time for
which control can be achieved. The agreement with the experiment is
reasonable, especially given that the very narrow tails of the domains
may be hard to detect in experiments.

We make the plausible assumption that the second largest Floquet mode
is a stable flip, so $\omega_2 = \pi /\tau$.  In our simple model,
there are then five parameters that determine the shapes of the
domains of control: $\lambda_2 \tau$ (the real part of the subleading
exponent), $A_1$ and $A_2$ (the amplitudes of the variation in
$\rho_1$ and $\rho_2$), and the phases $\phi_1$ and $\phi_2$ (which
determine where $\rho_1$ and $\rho_2$ vanish).

To fix these five parameters, we consider the $R=0.68$ domain.  The
phase $\phi_1$ must lie somewhere between $t_1\pi$ and $t_2\pi$ in
order for the divergence in $1/\rho$ and associated sign change in the
domain of control to be right.  From the fact that the onset of
divergence is not evident yet at $t_1$, we estimate that $\phi_1$ is
close to $t_2\pi$ and fix it at $0.6\pi$.  From the fact that the
boundary of the subleading mode does not appear to cut off the domain
near $t_l=0$, we take the divergence of $\rho_2$ to occur there,
requiring $\phi_2 = 0$ or $\pi$. The remaining parameters, $A_1$,
$A_2$, and $\lambda_2\tau$ are adjusted to fit $t_1$, $t_2$, the
latency times (in units of $\tau$) corresponding to the limits of the
domains, and $\gamma(t_2)$, the gain at which the lower domain of
control is cut off at $t_2$. The parameters determined from the
$R=0.68$ data are used for the $R=0$ plot as well since $\rho(t_l)$ is
determined purely from the uncontrolled system.

The primary conclusion we draw is that the theory does give
qualitative insight into the structure of the stability domains.  Even
with our crude constraints on the form of $\rho(t_l)$, the general
shapes of the domains are reproduced surprisingly well.  As expected,
larger $R$ increases the size of the domain of control, especially the
part with negative $\gamma$.

It may appear that by adjusting the full functions $\rho_1$ and
$\rho_2$ one could fit arbitrary shapes of the stability domains, and
it is true that many types of undulations in the domain boundaries
could be fit.  Moreover, one can always appeal to lack of robustness
to noise to explain why very narrow regions of a predicted stability
domain would not show up in the experimental data. There are, however,
some features of the experimental data that cannot be reproduced by
the theory presented here.  In particular, consider the width of the
region at negative $\gamma$ for $R=0.68$.  The upper and lower
boundaries of the right half of this region both represent
instabilities in the same mode, the mode associated with $\lambda_1$.
For a fixed value of $t_l$, then, the ratio of the $\gamma$'s at these
boundaries is independent of $\rho$.  The substantially larger width
of the experimental stability domain cannot be obtained by adjusting
any of the parameters in our theory.  Including additional modes or
making a different assumption about $\omega_2$ would not help, since
the narrowness of the theoretical domain is determined by $\lambda_1$.
We therefore conclude that second-order effects are significant in the
experimental system.  These effects may involve interactions between
modes associated with different $\lambda$'s or just interactions of
modes within the set generated by $\lambda_1$ alone.

\begin{figure}[t]
\begin{center}
\epsfxsize= 0.48 \linewidth
\epsfbox{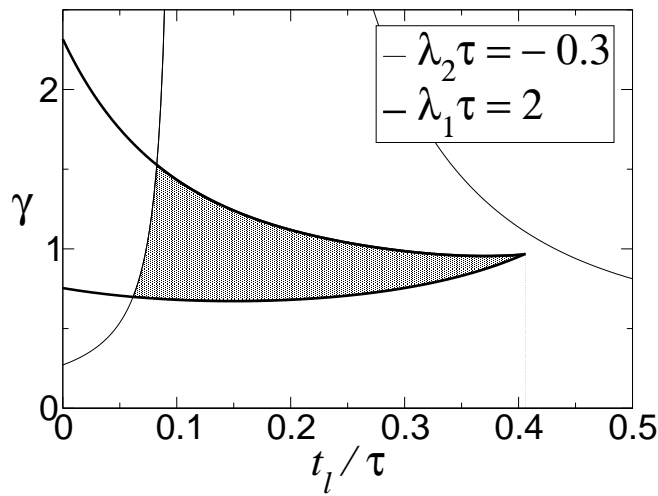}
\end{center}
\caption{\label{fig:non-zero}  
  Domain of control for two non-interacting Floquet modes, $\lambda_1
  \tau= 2$, $\lambda_2 \tau= -0.3$, and $\omega_{1} =\omega{2} =
  \pi/\tau$ for $R = 0.68$. The parameters $\phi$ and $A$ are chosen
  as $\phi_1 = 0.65 \pi,\: A_1 = 2.5$ and $\phi_2 = 0.1,\: A_2 = -3$.
  The shaded area is the effective domain of control.}
\end{figure}

Our analysis can also be used to explore the possibilities for
qualitatively different domain shapes.  An interesting example is
obtained when $\phi_2$, the phase associated with the subleading mode
of the uncontrolled system is taken to be $0.1\pi$ (and $A_2 < 0$).  As
shown in Figure~\ref{fig:non-zero}, this can lead to a situation in
which a nonzero latency time is {\em required} for effective control.

\section{Conclusion}
We have discussed the effects of latency time on a feedback control
scheme known as ETDAS. Using Floquet theory and carrying out a
first-order perturbation theory in the feedback gain, we have shown
that nontrivial domain shapes can arise in the plane parameterized by
feeback gain and latency time.  Within the first order theory, we find
that no control is possible above a maximum latency time determined
solely by the Floquet exponent of the most unstable mode in the
uncontrolled system.  We also find that Floquet modes that are stable
in the uncontrolled system contribute significantly to the overall
stability picture, reducing the domain of control substantially.

The theory accounts well for qualitative features of the stability
domains observed in experiments.  As expected, larger values of the
ETDAS parameter $R$ give larger stability domains.  Detailed
comparison indicates, however, that second order effects are
experimentally observable.

\begin{acknowledgments}
We would like to thank E. Sch\"oll, W. Just, A. Amann, I. Harrington, and
D. Gauthier for useful conversations. This work was supported by NSF
Grant PHY-98-70028 and within the framework of the exchange program
between TU Berlin and Duke University. P.H. acknowledges a Fulbright
Scholarship.
\end{acknowledgments}

%\bibliographystyle{/home/einstein/grad/hoevel/latex/prsty-fullauthor}
%\bibliography{/home/einstein/grad/hoevel/latex/ref}

\end{document}